\newif\ifpreprint\preprinttrue
\ifpreprint
\documentclass[prl,twocolumn,showkeys,amsfonts,letterpaper,floatfix]{revtex4}
\else
\documentclass[11pt]{article}
\fi

\def\DATE{July 14, 2004; revised September 9, 2004}

\usepackage{times}
\ifpreprint\else\usepackage{overcite}\usepackage{amssymb}\fi
\usepackage{graphicx}\paperwidth=8.5truein\paperheight=11truein
\usepackage[bookmarks=false]{hyperref}
\hypersetup{pdftitle={Method for Computing Protein Binding Affinity},
pdfauthor={C. Karney, J. Ferrara, S. Brunner}}

\ifpreprint\else
\fi

\ifpreprint\else
\fi

\newlength{\figwidth}
\ifpreprint
\else
\fi

\ifpreprint\else
\fi

\begin{document}

\title{Method for Computing Protein Binding Affinity}

\ifpreprint
\author{Charles F. F. Karney}\email{ckarney@sarnoff.com}
\author{Jason E. Ferrara}
\affiliation{\href{http://www.sarnoff.com}{Sarnoff Corporation},
  Princeton, NJ 08543-5300}
\author{Stephan Brunner}\altaffiliation
[Current address: ]{CRPP, Ecole Polytechnique F\'ed\'erale de Lausanne, 
CH-1015 Lausanne, Switzerland.}
\affiliation{\href{http://www.locuspharma.com}
 {Locus Pharmaceuticals, Inc.},
 Blue Bell, PA 19422-2700}
\fi

\date{\DATE}

\begin{abstract}
A Monte Carlo method is given to compute the binding affinity of a
ligand to a protein.  The method involves extending configuration space
by a discrete variable indicating whether the ligand is bound to the
protein and a special Monte Carlo move which allows transitions between
the unbound and bound states.  Provided that an accurate protein
structure is given, that the protein-ligand binding site is known, and
that an accurate chemical force field together with a continuum
solvation model is used, this method provides a quantitative estimate of
the free energy of binding.
\ifpreprint
\fi
\end{abstract}

\ifpreprint \maketitle \fi

\section*{Introduction}

Many drugs work by binding to a target protein in an organism and
affecting the action of this protein.  The binding of the drug molecule,
the ligand $\mathrm L$, to the protein $\mathrm P$ under physiological
conditions is usually reversible (characterized by weak chemical
interactions rather than covalent bonds),
\[
\mathrm L + \mathrm P \rightleftharpoons \mathrm{LP},
\]
and, in equilibrium and in the dilute
limit, the concentration of the ligand-protein complex $[\mathrm{LP}]$
is given by the dissociation constant
\[
K_d = \frac{[\mathrm L][\mathrm P]}{[\mathrm{LP}]}.
\]
It is convenient to define the binding affinity as
\[\mathrm pK_d =
-\log_{10}\biggl(\frac{K_d/N_\mathrm A}{1\,\mathrm{kmol\,m^{-3}}}\biggr),
\]
where $N_\mathrm A$ is the Avogadro constant.  A high value for
$\mathrm pK_d$ is crucial to obtaining a good drug molecule and,
consequently, the ability to compute $\mathrm pK_d$ accurately would
greatly accelerate drug discovery by allowing many molecules to be
screened \emph{in silico} before any time-consuming syntheses and assays
are done.  The dissociation constant can also be expressed in terms of
the binding free energy $\Delta F$,
\begin{equation}
\label{kdfree}
K_d = \exp(\beta \Delta F)/V_0,
\end{equation}
where $V_0$ is the system volume, $\beta = 1/(kT)$, $T$ is the
temperature, and $k$ is the Boltzmann constant.  Similarly $\mathrm
pK_d$ is given by
\[
\mathrm pK_d = -\frac{\beta \Delta F}{\ln10} +
\log_{10}(V_0 N_\mathrm A\times 1\, \mathrm{kmol\, m^{-3}}).
\]
The quantity $\Delta F$ is the free energy difference of the ligand and
the protein forming a bound complex $\mathrm{LP}$, the ``bound system,''
compared to the ligand and the protein isolated from one another
$\mathrm L + \mathrm P$, the ``unbound system.''

In order to compute $\mathrm pK_d$, we require (1)~a sufficiently
accurate model of the protein and the ligand and their interaction and
(2)~a good way to compute the resulting value of $\mathrm pK_d$.  In
this paper, we assume the first requirement is fulfilled and instead
focus on meeting the second.

The standard methods of computing free energies
\nocite{mezei86}\cite{mezei86,kollman93}
are not capable of computing $\Delta F$ directly because the unbound and
bound systems are too dissimilar, which hinders transitions between
these systems.  Instead, typically, two close ligands $\mathrm L_a$ and
$\mathrm L_b$, are compared separately unbound and bound to the protein,
thereby obtaining the difference in the free energies $\Delta F_a -
\Delta F_b$.

We present here a practical method for directly computing $\Delta F$ and
hence $\mathrm pK_d$.  The method consists of: (1)~formulating the
problem in an extended phase space which allows the unbound and bound
systems to be treated as a single system and $K_d$ to be expressed as
the ratio of two canonical averages; (2)~introducing a new Monte Carlo
move, the ``wormhole move,'' to
make transitions between the unbound and bound states in this extended
system; and (3)~a method to determine the ``portals'' needed for
the wormhole move.

\section*{Formulation}

\ifpreprint
Consider a system of volume $V_0$ consisting of a ligand molecule
$\mathrm L$ and a protein molecule $\mathrm P$ dissolved in $N_\mathrm
S$ molecules $\mathrm S$.  The overall state of the system is given by
$[\Gamma, \Gamma_\mathrm S]$ where $\Gamma$ represents the phase space
configuration of $\mathrm L$ and $\mathrm P$ and $\Gamma_\mathrm S$
represents the configurations of all the solvent molecules $\mathrm S$.
The energy of the system is given by $E(\Gamma, \Gamma_\mathrm S)$
and, in equilibrium, the system obeys the Boltzmann distribution
\cite{landau69}
\[
f(\Gamma, \Gamma_S) = \frac
{\exp[-\beta E(\Gamma, \Gamma_S)]}
{\int \exp[-\beta E(\Gamma, \Gamma_S)]\, d\Gamma\,d\Gamma_\mathrm S}
.
\]
It is frequently useful to average over the configurations of the
solvent molecules by integrating the Boltzmann distribution over
$\Gamma_\mathrm S$ to give a reduced Boltzmann distribution
\begin{eqnarray*}
f(\Gamma) &=&
\int f(\Gamma, \Gamma_\mathrm S) \,d\Gamma_\mathrm S\\
&=& \frac
{\exp[-\beta E(\Gamma)]}
{\int \exp[-\beta E(\Gamma)]\, d\Gamma}
,
\end{eqnarray*}
where
\[
E(\Gamma) = -\frac1\beta \ln\biggl(
\int\exp[-\beta E(\Gamma, \Gamma_\mathrm S)]\,d\Gamma_\mathrm S
\biggr)
\]
is the energy of the system with ligand and protein configurations
specified by $\Gamma$ and with the equilibrium effects of the solvent
implicitly included as a solvation free energy.
In this paper, we will assume that $E(\Gamma)$ is
given.

Typical molecular interactions have a short range.  In view of this, let
us define $\Sigma_0$ to represent all accessible $\Gamma$ space
(i.e., $\mathrm L$ and $\mathrm P$ somewhere in the system volume
$V_0$), and $\Sigma_1$ to represent that portion of $\Sigma_0$
where there is an appreciable interaction between $\mathrm L$ and
$\mathrm P$ which therefore form the complex $\mathrm{LP}$.  In the
phase-space volume $\Sigma_1$ we write the energy as $E_1(\Gamma)$
which is just an alternate notation for the full energy $E(\Gamma)$,
while in the volume $\Sigma_0 - \Sigma_1$ we may write the
energy as $E_0(\Gamma)$ which we define as the ``unbound'' energy, i.e.,
$E(\Gamma)$ excluding the interaction between $\mathrm L$ and $\mathrm
P$.  The dissociation constant can then be written as
\begin{eqnarray*}
K_d &=& \frac1{V_0}\frac
{\Bigl(
\int_{\Sigma_0 - \Sigma_1} e^{-\beta E(\Gamma)} \,d\Gamma
\Bigr)^2}
{
\int_{\Sigma_0} e^{-\beta E(\Gamma)} \,d\Gamma
\int_{\Sigma_1} e^{-\beta E(\Gamma)} \,d\Gamma}\\
&=& \frac1{V_0}\frac
{\Bigl(
\int_{\Sigma_0 - \Sigma_1} e^{-\beta E_0(\Gamma)} \,d\Gamma
\Bigr)^2}
{\int_{\Sigma_1} e^{-\beta E_1(\Gamma)} \,d\Gamma}\times\\
&&\qquad
\frac1
{\int_{\Sigma_0 - \Sigma_1} e^{-\beta E_0(\Gamma)} \,d\Gamma
+ \int_{\Sigma_1} e^{-\beta E_1(\Gamma)} \,d\Gamma}
.
\end{eqnarray*}
In the dilute limit $V_0 \rightarrow \infty$, this can be simplified
by ignoring the second term in the denominator of the last factor and by
extending the limits of the integrals over $\Sigma_0 - \Sigma_1$
to include $\Sigma_1$.  In extending the definition of $E_0(\Gamma)$
to $\Gamma\in\Sigma_1$, we include the intramolecular energy and the
solvation free energy but continue to omit the intermolecular
(ligand-protein) energy.  This yields
\nocite{bennett76}\cite{mezei86,bennett76,luo02}
\begin{equation}\label{kd}
K_d =  \frac1{V_0}\frac
{\int_{\Sigma_0} \exp[-\beta E_0(\Gamma)] \,d\Gamma}
{\int_{\Sigma_1} \exp[-\beta E_1(\Gamma)] \,d\Gamma}
.
\end{equation}
\fi
The Helmholtz free energy of the unbound and bound
systems is \cite{landau69}
\[
F_\lambda = -\frac1\beta\ln\biggl(
  \int_{\Sigma_\lambda} \exp[-\beta E_\lambda(\Gamma)] \,d\Gamma
\biggr),
\]
for $\lambda = 0$ and $1$, and Eq.~(\ref{kdfree}) is obtained from
Eq.~(\ref{kd}) with $\Delta F = F_1 - F_0$.  The definition of $K_d$,
Eq.~(\ref{kd}), is strictly independent of $V_0$ because of
translational symmetry (ignoring boundary effects).  It is also
independent of the precise definition of $\Sigma_1$, provided that
$\Sigma_1$ includes the protein-ligand binding site and does not
include a ``macroscopic'' volume beyond this.

In this formulation, we have assumed that the system volume $V_0$ is
fixed.  However, in most physiological systems, the pressure is held
constant and the binding affinity is then related to the differences in
the Gibbs free energy which introduces a correction term which is the
product of the pressure and the change in the volume caused by the
formation of the $\mathrm{LP}$ complex \cite{gilson97}.  We expect this
correction to be small for typical ligand-protein interactions in
solution.

We would like to cast Eq.~(\ref{kd}) as the ratio of canonical averages
which can be computed using the canonical-ensemble Monte Carlo method
\cite{metropolis53}.  To achieve this, we combine the unbound and bound
systems by extending phase space to include a discrete index $\lambda\in
\{0,1\}$ and consider a system in this extended space with energy
$E^*_\lambda(\Gamma)$ for which the canonical average is defined by
\[
\langle X \rangle = \frac
{\sum_\lambda \int d\Gamma\, \exp[-\beta E^*_\lambda(\Gamma)]
  X_\lambda(\Gamma)}
{\sum_\lambda \int d\Gamma\, \exp[-\beta E^*_\lambda(\Gamma)]}
.
\]
We take $E^*_\lambda(\Gamma)$ to be infinite for $\Gamma \notin
\Sigma_\lambda$ and finite otherwise.  Now Eq.~(\ref{kd}) can be
rewritten as
\begin{equation}\label{kdcanon}
K_d = \frac1{V_0} \frac
{\bigl< \delta_{\lambda0} \,
  \exp\bigl(-\beta [E_0(\Gamma) - E^*_0(\Gamma)]\bigr) \bigr>}
{\bigl< \delta_{\lambda1} \,
  \exp\bigl(-\beta [E_1(\Gamma) - E^*_1(\Gamma)]\bigr) \bigr>}
,
\end{equation}
where $\delta_{\lambda\mu}$ is the Kronecker delta.  If
$E^*_\lambda(\Gamma) \approx E_\lambda(\Gamma)$, the terms being
averaged are $O(1)$.  Because the definition of $K_d$ is independent of
$V_0$, we can pick $V_0 \sim 1/K_d$ so that approximately the same
number of samples contribute to each of the canonical averages.
We show later, Eq.~\ref{err}, that this choice minimizes the error in
the estimate of $K_d$.

We can evaluate $E^*_\lambda(\Gamma)$ with short energy cutoffs allowing
it to be computed more quickly than $E_\lambda(\Gamma)$ and the terms
contributing to the averages in Eq.~(\ref{kdcanon}) can be accumulated
every hundred steps, for example.  Since there is typically a high
correlation between successive steps in a Monte Carlo simulation, this
method allows the averages to be computed to a given degree of accuracy
in less time than if we had used $E^*_\lambda(\Gamma) = E_\lambda(\Gamma)$.

The extension of phase space has been used in other free energy
calculations, to combine, for example, systems at several different
temperatures \cite{lyubartsev92} or to treat the ``reaction coordinate''
controlling the transition between two chemical species as a dynamic
variable \nocite{tidor93}\cite{tidor93,kong96}.
In our case, the use of the wormhole
Monte Carlo (described in the next section) allows us to include just the two
systems of interest without the need to compute the properties of
(possibly unphysical) intermediate systems.

\section*{Wormhole Monte Carlo}

\begin{figure}
\begin{center}%
\includegraphics[width=0.8\figwidth]{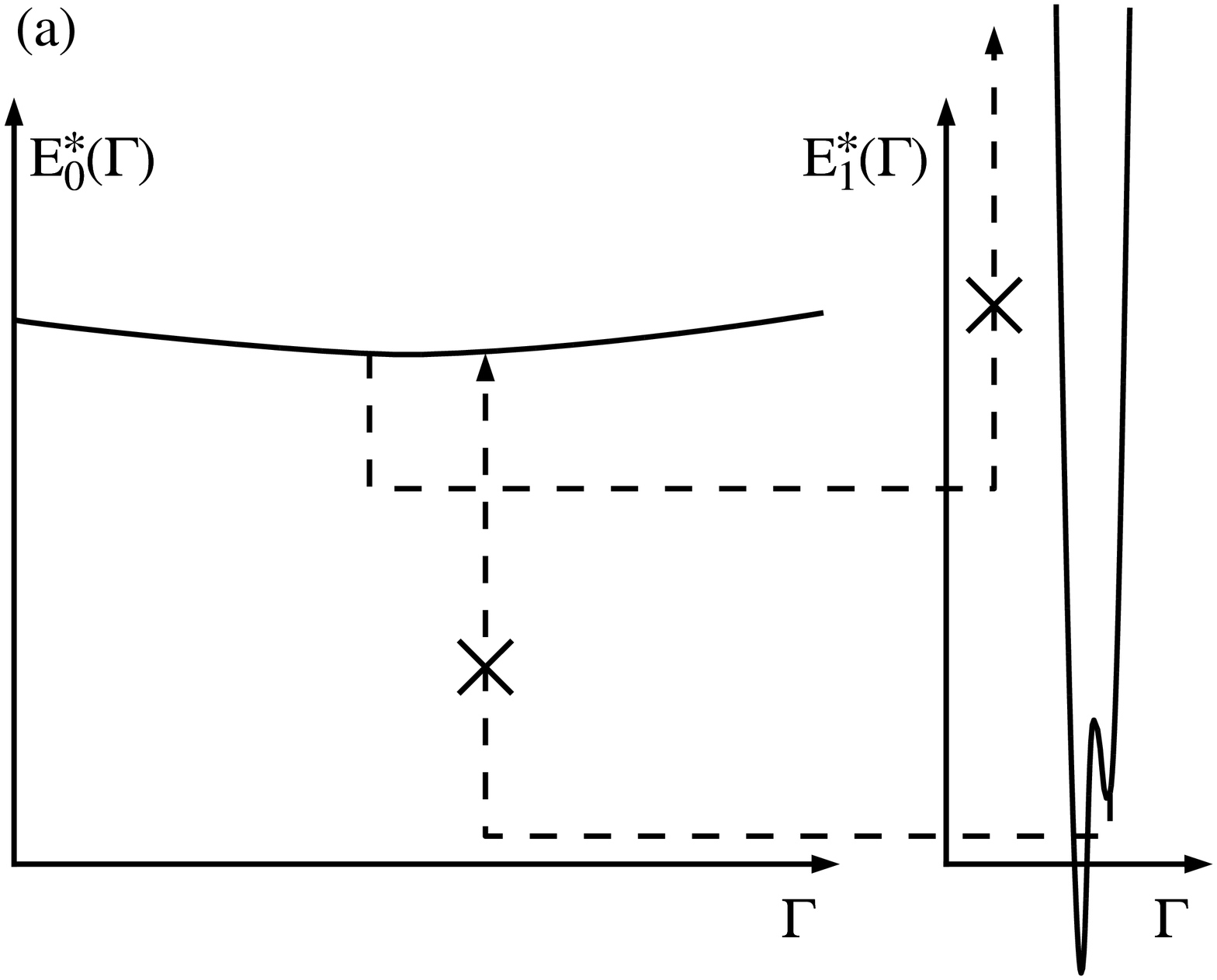}\\[0.5ex]
\includegraphics[width=0.8\figwidth]{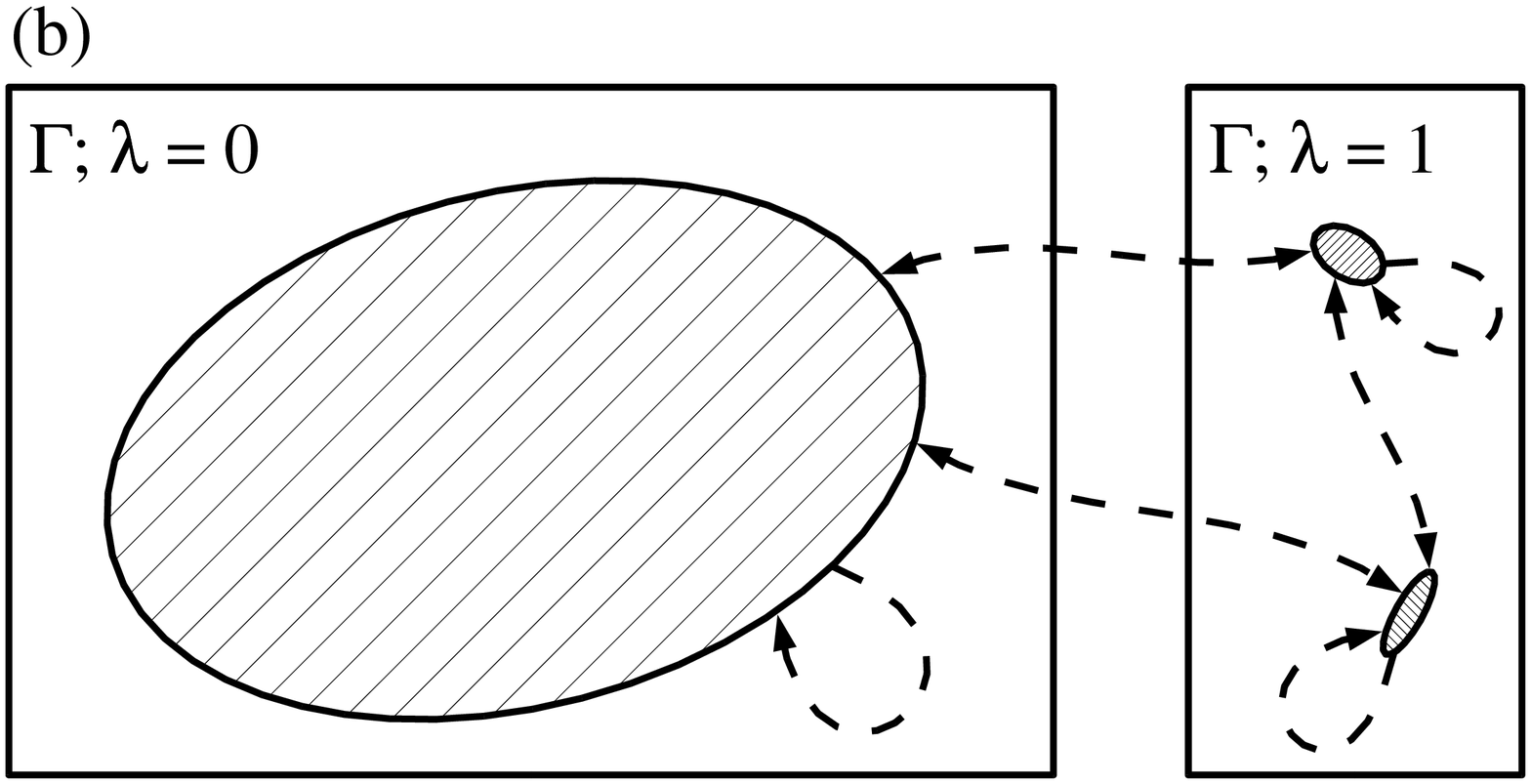}\\[-1.5ex]
\end{center}
\caption{\label{energyfig}
(a) Schematic representation of $E^*_0(\Gamma)$ (shallow and wide)
and $E^*_1(\Gamma)$ (deep and narrow).  Conventional Monte Carlo moves
between $\lambda = 0$ and $1$ (shown as dashed lines) are nearly always
rejected because they lead to large increases in energy.
(b) Schematic representation of typical portals for the wormhole moves
for the case illustrated in (a),  The large ratio of the volume of the
unbound ($\lambda = 0$) portal compared to the bound ($\lambda = 1$)
portals compensates for the higher energy of the unbound configurations.
This results in accepted wormhole moves (dashed lines) between all the
portals.
}
\end{figure}

We can compute the canonical averages in Eq.~(\ref{kdcanon}) using
the Monte Carlo method \cite{metropolis53} to make steps from $[\Gamma,
\lambda]$ to $[\Gamma', \lambda']$ with probability
\[
\min\bigl[1,
  \exp\bigl(-\beta [E^*_{\lambda'}(\Gamma')- E^*_\lambda(\Gamma)]\bigr)
\bigr].
\]
However, the estimate of the ratio in Eq.~(\ref{kdcanon}) will be very
poor, because transitions between $\lambda = 0$ and $1$ will be
extremely rare---typically, $E^*_0$ is shallow and wide, while $E^*_1$
is deep and narrow; see Fig.~\ref{energyfig}(a).
One possible way of remedying this is to treat $\lambda$ as a continuous
variable \cite{tidor93,kong96}, providing a suitably interpolated
definition of $E_\lambda(\Gamma)$, and allowing Monte Carlo steps with
small changes in $\lambda$.  In practice, this approach only ``works''
if the two endpoints are sufficiently similar, thus limiting this method
to the comparison of chemically close molecules.

Here we propose an alternative way of carrying out a Monte Carlo
simulation of the $E^*_\lambda(\Gamma)$ system.
We restrict the standard
moves to changes in $\Gamma$ only, and allow changes in $\lambda$ via
``wormhole moves'' \cite{whnote} which connect otherwise disconnected
regions of configuration space.
This obviates the
need to treat (possibly unphysical) intermediate values of $\lambda$,
 permitting us to compute the free energy differences needed to
determine the absolute binding affinity.

Assume we have some system defined on a phase space $\Upsilon$ whose
equilibrium distribution is proportional to $g(\Upsilon)$.  The
canonical average of a quantity $X(\Upsilon)$ is defined by
\[
\langle X\rangle = \frac
{\int d\Upsilon\,g(\Upsilon) X(\Upsilon)}
{\int d\Upsilon\,g(\Upsilon)}.
\]
In our application, we make the identification $\Upsilon =
[\Gamma,\lambda]$, $\int d\Upsilon = \sum_\lambda \int d\Gamma$, and
$g(\Upsilon)=\exp[-\beta E_\lambda(\Gamma)]$.

Let us define a set of ``portal functions,'' $w$, $w'$, $w''$, \ldots,
on $\Upsilon$, with properties
\[
0\le w(\Upsilon) \le 1/v < \infty,
\]
\[
\int d\Upsilon\, w(\Upsilon) = 1,
\]
where $v$ is a representative phase-space volume of the portal
function.  A wormhole move consists of the following steps: select a
pair of portals $(w, w')$ with probability $p_{ww'}$; reject the move
with probability $1 - v w(\Upsilon)$, where $\Upsilon$ is the current
state; otherwise, with probability $v w(\Upsilon)$, pick a
configuration $\Upsilon'$ with probability $w'(\Upsilon')$; and accept
the move to $\Upsilon'$ with probability $P_{ww'}(\Upsilon,\Upsilon')$.
If the move is rejected, $\Upsilon$ is retained as the new state.
We determine
$P_{ww'}(\Upsilon,\Upsilon')$ by demanding that the rate of making
transitions from $\Upsilon$ to $\Upsilon'$ via portals $(w,w')$ is
balanced by the reverse rate from $\Upsilon'$ to $\Upsilon$ via
portals $(w',w)$, i.e.,
\[
R_{ww'}(\Upsilon, \Upsilon') = R_{w'w}(\Upsilon', \Upsilon)
\]
(the condition of detailed balance), where the rates are given by
\begin{eqnarray*}
R_{ww'}(\Upsilon, \Upsilon') &=&
p_{ww'} \frac{g(\Upsilon)}
{\int d\Upsilon\,g(\Upsilon)}\times\\
&&\qquad
v w(\Upsilon) w'(\Upsilon') P_{ww'}(\Upsilon,\Upsilon').
\end{eqnarray*}
A possible solution for the acceptance probability is
\[
P_{ww'}(\Upsilon,\Upsilon') =
\min\biggl(1, \frac{p_{w'w}}{p_{ww'}} \frac{g(\Upsilon')}{g(\Upsilon)}
\frac{v'}{v}\biggr),
\]
where we have made use of the identity $\alpha \min(1, \alpha^{-1}) =
\min(1,\alpha)$, which is valid for $\alpha > 0$.

In order to apply this move, let us use a specific rather simple form
for the portal functions, $w(\Upsilon)$, namely
\[
w(\Upsilon) = \left\{
\begin{array}{l@{\hspace{1em}}l}
1/v, & \mbox{for $\Upsilon \in w$},\\
0, & \mbox{otherwise},
\end{array}\right.
\]
where we now denote a portal by $w$ which defines an arbitrary subset of
$\Upsilon$ space of volume $v$.  In principle, there is no restriction on the
choice of the portals; however, practical considerations, discussed
below, dictate how they are chosen.  Furthermore, for simplicity, we
will assume that the wormhole probabilities are all equal,
$p_{ww'} = \mathrm{const}$.

Let us now describe the wormhole move in $[\Gamma,\lambda]$ space.
Starting with a configuration $[\Gamma, \lambda]$, first pick a
random portal $w$.  If $[\Gamma, \lambda] \notin w$, then reject the
move.  Otherwise, pick a random configuration $[\Gamma', \lambda']$
uniformly in a randomly chosen portal $w'$, and accept the move to
$[\Gamma',
\lambda']$ with probability
\begin{equation}\label{whacc}
\min\biggl(1,
\frac{\exp[-\beta E^*_{\lambda'}(\Gamma')]}
{\exp[-\beta E^*_\lambda(\Gamma)]}
\frac{v'}v
\biggr).
\end{equation}
This differs from the standard Boltzmann acceptance probabilty by
the ratio of volumes, $v'/v$, which can be large enough to compensate for
the difference in the mean energies in the portals.  Indeed, if the
mean energy of configurations in $w$ scales as
$\beta^{-1} \ln v + \mathrm{const.}$, the acceptance probability,
Eq.~(\ref{whacc}), is $O(1)$, thereby allowing moves between
shallow wide wells and deep narrow ones; see Fig.~\ref{energyfig}(b).
In practice, each portal will occupy one of the energy wells of either
$E^*_0(\Gamma)$ or $E^*_1(\Gamma)$.  This implies that the $\lambda = 0$
portals will permit unrestricted translation and rotation of the
ligand and the protein subject to the $V_0$ constraint, and the $\lambda
= 1$ portals will allow unrestricted motion of one of the molecules.

The wormhole move embodies two concepts which have been used separately
in other works: stretching or shrinking $\Gamma$ space when making the
move compensates for possibly large energy differences between wells
\nocite{miller00}\cite{miller00,zhu02};
and jumping between disconnected regions of phase
space enables the Markov chain to explore regions of phase space
separated by large energy barriers
\nocite{voter85}\cite{voter85,senderowitz95}.  In the
following sections, we will show how these elements may be combined to
permit the computation of protein binding affinities with realistic
force fields.

\section*{Finding the portals}

In order for the wormhole method to be practical, we need a reliable way
of choosing the portals.  We describe this process first in the
general case.  Let us write $\Gamma = [\Gamma_\mathrm L, \Gamma_\mathrm
P]$, where $\Gamma_\mathrm M = [X_\mathrm M, \Xi_\mathrm M]$ represents
the state of molecule $\mathrm M$, $X_\mathrm M$ represents its position
and orientation, and $\Xi_\mathrm M$ represents its conformation.

We assume that $\Xi_\mathrm M$ is expressed in such a way that any
constraints on the positions of the atoms in $\mathrm M$ (e.g., bond
lengths and bond angles) is implicitly accounted for, so that the
dimensionality of $\Xi_\mathrm M$ reflects the number of degrees of
conformational freedom, $n_\mathrm M$, for this molecule.

Because of the translational and rotational symmetry of the system,
certain components of $\Gamma$ are ignorable.  We can therefore write
$E^*_\lambda(\Gamma) = E^*_\lambda(\Xi_\lambda)$, with
\begin{eqnarray*}
\Xi_0 &=& [\Xi_\mathrm L,\Xi_\mathrm P],\\
\Xi_1 &=& [Y, \Xi_\mathrm L,\Xi_\mathrm P],
\end{eqnarray*}
where $Y = X_\mathrm L - X_\mathrm P$ denotes the position and
orientation of the ligand relative to the protein.  The dimensionality
of $\Xi_\lambda$ is $n_\lambda$ with $n_0 = n_\mathrm L + n_\mathrm P$
and $n_1 = n_0 + 6$.

The strategy for determining the portals is to carry out conventional
canonical Monte Carlo simulations with $E^*_\lambda(\Xi_\lambda)$
separately for $\lambda = 0$ and $1$.  For each $\lambda$, we obtain a
canonical set of configurations $\{\Xi\}$ (suppressing the $\lambda$
subscripts for brevity), to which we fit $n$-dimensional ellipsoids, as
follows.  First we try to fit a single ellipsoid to $\{\Xi\}$.  The
center of the ellipsoid is given by the mean configuration
$\langle\Xi\rangle$.  For each configuration in $\{\Xi\}$, we determine
the deviation from the mean, $\delta \Xi = \Xi - \langle\Xi\rangle$, and
compute a covariance matrix for the configurations which can be
diagonalized as
\[
\langle \delta\Xi \, \delta\Xi\rangle =
\mathsf P \Lambda \mathsf P^\mathrm T,
\]
where $\mathsf P$ is the matrix of (column) eigenvectors and $\Lambda$
is the diagonal matrix of eigenvalues.  Because of the properties of the
covariance, $\mathsf P$ is real and orthogonal, $\mathsf P^{-1} =
\mathsf P^\mathrm T$, and the eigenvalues are real and non-negative.  If there
are no hidden constraints on the motion, we additionally can assume that
the eigenvalues are strictly positive.  We find it convenient to define
\[
\mathsf B = \mathsf P \Lambda^{1/2}, \qquad
\mathsf B^{-1} =  \Lambda^{-1/2} \mathsf P^\mathrm T,
\]
so that we can write
\[
\langle \delta\Xi \, \delta\Xi\rangle = \mathsf B \mathsf B^\mathrm T.
\]
We take the semi-axes of the ellipsoid to be the columns of $\sqrt{n}
\mathsf B$.  The multiplier here, $\sqrt{n}$, is chosen to ensure that
$O(1)$ of the configurations in $\{\Xi\}$ lie within the
ellipsoid.  This choice is motivated by considering a symmetric
$n$-dimensional Gaussian
\[
f(\mathbf r) = \frac{\exp(-\frac12 r^2)}{(2\pi)^{n/2}},
\]
for which we have
\[
\langle r^2 \rangle = \int r^2 f(\mathbf r)\, d^n\mathbf r = n.
\]

Ellipsoids are a natural choice to use to fit the set of configurations
for the following reasons: (1)~The iso-density contours of the
distribution in a harmonic well are ellipsoids.  (2)~It is easy to
sample points randomly from an ellipsoid.  (3)~Conversely, it is easy to
test that a point lies inside an ellipsoid.  (4)~The volume of an
$n$-dimensional ellipsoid is given by
\[
v_n = \frac{\pi^{n/2}}{(n/2)!} \prod_{i=1}^n a_i,
\]
where $a_i$ is the length of $i$th semi-axis.  Note well the degenerate
case of this result, $v_0 = 1$, which is used if the protein and ligand
are both rigid ($n_0 = 0$).  The sampling and testing of points, (2) and
(3), can either be accomplished by transforming with the matrix $\mathsf
B$.  Alternatively, we can use the simpler Cholesky decomposition of the
covariance matrix
\[
\langle \delta\Xi \, \delta\Xi\rangle =
\mathsf C \mathsf C^\mathrm T,
\]
where $\mathsf C$ is a lower triangular matrix.  Both $\mathsf C$ and
$\mathsf C^{-1}$ may be computed by direct (non-iterative) methods
\cite{golub96}.

Having defined an ellipsoid in this way, we test its suitability as a
portal by demanding that $O(1)$ of the configurations sampled
uniformly from it have energies close to its mean energy $\langle
E^*(\Xi)\rangle$.  If this test fails, we split $\{\Xi\}$ into two sets
according to the sign of $\delta \Xi$ projected along the largest
semi-axis of the ellipsoid and construct new trial portals from each
of these sets.

The ellipsoids that result from this process constitute our portals.
When computing the volumes of the portals for use in
Eq.~(\ref{whacc}), we need to account for the freedom to place
$2-\lambda$ molecules at arbitrary positions and orientations in the
volume $V_0$.  In practice, this means we multiply the volume of the
$\lambda=0$ portals by $\sigma V_0$ where $V_0$ is the translational
volume and $\sigma$ is the orientational volume (given below).

In order to complete the specification of the portals, we need to
describe how $\langle\Xi\rangle$ and $\delta\Xi$ are formed, since, as a
consequence of working in the subspace where the molecular constraints
are implicitly satisfied, $\Xi$ is not simply a point in $\mathbb R^n$.
We wish to represent each ellipsoid on a locally Cartesian space
$\mathbb R^n$ which lets us use familiar formulas for defining the
ellipsoid. We also demand that the mapping to $\mathbb R^n$ have constant
Jacobian in order to maintain detailed balance when sampling from the
ellipsoids.

We illustrate this by considering the case of the protein and the ligand
being made up of $n_\mathrm M+1$ rigid fragments simply connected by
$n_\mathrm M$ bonds each of which allow rotation only (i.e., the bond
lengths and bond angles are fixed).  The relative position and
orientation of $\mathrm L$ with respect to $\mathrm P$, $Y$, can be
defined in terms of one of the rigid fragments of each molecule.
Finally, we use unit quaternions
\cite{hamilton47} to represent orientation.  Since the quaternions
$q$ and $-q$
represent the same rotation, orientations are given by an \emph{axis}
\cite{mardia99} of the sphere $S^3$.

The coordinates making up $\Xi$ are then: (a)~the position component of
$Y$, a point in $\mathbb R^3$; (b)~the orientation component of $Y$, an
axis of the sphere $S^3$; and (c)~the dihedral angles of the rotatable
bonds of the ligand and protein, points on the circle $S^1$.  The
definition of
$\langle\Xi\rangle$ and $\delta\Xi$ is straightforward for (a), since we
use the normal arithmetic definitions.  For (c), we define the mean for
each angle \cite{mardia99} by the direction of the mean of the points on
$S^1$ embedded in $\mathbb R^2$.  We form the deviation in this case by
subtraction modulo $2\pi$ so that the result lies in $[-\pi,\pi]$.

To find the mean of the orientations (b), we similarly embed $S^3$ in
$\mathbb R^4$.  We define the mean orientation \cite{mardia99} as the
axis in $\mathbb R^4$ about which the moment of inertia of the sample
axes is minimum.
To find the deviation of the orientation, we compute
the differential rotation $\delta q = q \langle q\rangle^*$ which takes
the mean orientation to the sample orientation and we project this into a
``turn'' vector $\mathbf u$ in the unit ball in $\mathbb R^3$ so as to
preserve the metric.  This is achieved by a generalization of the
Lambert azimuthal equal-area projection as described in the Appendix.
In this representation, the volume of orientational
space (which contributes to the multiplier for the unbound volumes) is
$\sigma = \frac43\pi$.

Occasionally, the point sampled from $w'$ corresponds to one of the
coordinates ``wrapping around,'' i.e., the change in the dihedral lies
outside $[-\pi,\pi]$ or the turn $\mathbf u$ lies outside the unit ball.
In order to preserve detailed balance, we reject the resulting move.

There are four ways in which we can improve the quality of the
portals obtained by this method so as to make successful wormhole
moves more likely.  (1)~The Monte Carlo runs used to obtain the
samples from which the portals are defined should begin with an
``annealing'' phase where the temperature is started at some high value
and slowly reduced to $T$ at which point we start gathering samples.
This allows the Monte Carlo sampling to explore phase space more
thoroughly.
(2)~Other methods of finding conformational energy minima \cite{head97}
can be applied to provide additional starting points for the Monte Carlo
runs.
(3)~The samples can be supplemented with those
obtained by applying those symmetry operations which leave the molecules
invariant.  In the case of non-chiral ligands, we would also apply a
reflection of the ligand in the unbound case since this will leave the
energy unchanged.  In this way, the portal moves allow all the
symmetric variants of the molecules to be explored so that symmetry is
included in a systematic way in the computation of the binding affinity.
(4)~When forming $\langle \delta\Xi_0 \, \delta\Xi_0\rangle$, we
should set the intermolecular cross terms to zero because the
conformations of the two molecules are independent when $\lambda = 0$.

This method of finding portals depends on the samples ``spanning'' a
volume of phase space.  This requires that the dimensionality of phase
space be sufficiently small and this, in turn, implies the use of an
implicit solvation model.  In addition the portals can be more
reliably found if the ``hard'' degrees of freedom in the molecules are
replaced by constraints (e.g., by fixing the bond lengths and bond
angles as described above).

The method of successively subdividing the samples may lead
to suboptimal portals in some cases, for example, by dividing a
contiguous set of samples.  We have recently experimented with fitting a
mixture of Gaussians to the samples using the expectation-maximization
(EM) algorithm \nocite{dempster77}\cite{dempster77,verbeek03}.
Since this optimizes the
fit to all the samples, it usually results in fewer, better, portals.
In this case, we are naturally lead to use Gaussians for the portal
functions rather than the more restrictive ellipsoids.  By enforcing the
symmetries of the ligand when making the fits, ligand symmetry can also
be included in a rigorous way.  These improvements will be described in a
subsequent publication.

\section*{The free energy calculation}

Prior to the calculation of the free energy, we estimate a suitable
value of $V_0$, which enters into the definition of the volumes of the
$\lambda=0$ portals, by assigning an estimated statistical weight of
$v\exp(-\beta \langle E^*\rangle)$ to each portal and estimating
\[
V_0 \sim \frac
{\sum_{\lambda=1} v \exp(-\beta \langle E^*\rangle)}
{\sum_{\lambda=0} (v/V_0) \exp(-\beta \langle E^*\rangle)}
,
\]
where the sums in numerator (denominator) are over the bound (unbound)
portals and $v/V_0$ for the unbound portals is the volume of the
ellipsoid (multiplied by $\sigma$) and thus does \emph{not} depend on
$V_0$.
If this estimate for $V_0$ results in the sampling being
too heavily weighted toward $\lambda =
0$ or $1$, $V_0$ may be changed and
the current values of the sample sums for $K_d$ can be
adjusted to account for this change.

The choice of the starting configuration for the free energy calculation
may introduce some bias in the results.  We can remove much of this bias
by using the portals to select the starting configuration:
select a portal $w$ with probability
proportional to its estimated statistical weight; select a
configuration, $[\Gamma, \lambda]$, uniformly from this portal; and
accept the configuration with probability
\[
\min\bigl[1, \exp\bigl(-\beta[E^*_\lambda(\Gamma) -
  \langle E^*\rangle]\bigr)\bigl],
\]
where $\langle E^*\rangle$ is the canonical average energy over the
portal.  This procedure is repeated until a configuration is accepted.
The bias can be further reduced by running the Monte Carlo calculation
for several correlation times, defined by Eq.~(\ref{diffusion}), prior to
gathering the data for Eq.~(\ref{kdcanon}).

During the course of the free energy calculation, normal and wormhole
Monte Carlo moves are mixed.  With the normal moves, we only attempt to
change $\Gamma$ and, for this reason, it is possible to have the Monte
Carlo step size depend on $\lambda$ (with, usually, the step size being
larger with $\lambda = 0$).
In practice, most ($\sim 90\%$) of the attempted moves are wormhole moves
because frequently we have
$[\Gamma,\lambda] \notin w$ and the attempted wormhole move is inexpensively
rejected.

The method is robust in the sense that it does not depend on the
particular form of the energy function.  In addition, a failure to make
transitions between $\lambda = 0$ and $1$ can be detected.  This may be
because the test $[\Gamma, \lambda] \in w$ never succeeds (i.e., the
configuration has been trapped in a new well), or because the acceptance
criterion is never met, which indicates that there is a deep well within
the well of one of the portals.  In both cases, the problem can be
corrected by adding a new portal based on recent configurations.

\section*{Example}

The efficacy of the wormhole method depends on how frequently a
configuration lies within one of the portals and how often the jump to
the new portal is accepted.
\begin{figure}
\begin{center}
\vspace{1.0ex}
\includegraphics[height=9ex]{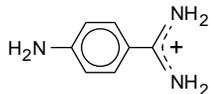}
\vspace{-1.5ex}
\end{center}
\caption{\label{benzamidine}
The structure of $p$-amino-benzamidine.
}
\end{figure}
In order to assess these questions, we
have computed the binding affinity of $p$-amino-benzamidine, whose
structure is shown in Fig.~\ref{benzamidine},
to the digestive enzyme trypsin, at $T = 290\,\mathrm K$.
We emphasize that the primary goal of this exercise is to assess how
well wormhole moves allow the free energy calculation to converge.  For
this purpose, we are not so interested in comparing the resulting
computed binding affinity to the experimental data since this will
depend in large measure on the accuracy of the force field and of the
protein structure.  Nevertheless, since the convergence will depend on
the complexity of the energy ``landscape,'' we use this example to
epitomize the binding of a small ligand to a protein.

The coordinates of the atoms in
trypsin are taken from a trypsin-benzamidine complex, 1BTY
\cite{katz95}.  At physiological $\mathrm{pH}$, the amidine group is
protonated (net charge of $+1$) and, in the complex, it is attracted to
a negatively charged aspartate residue in trypsin inhibiting its
enzymatic action.  We employ the Amber 7 force field
\nocite{amber7}\nocite{cornell95}\cite{amber7,cornell95,bayley93}
and the GB/SA solvation model
\nocite{still90}\nocite{hawkins95}\cite{still90,hawkins95,tsui00}.
The protein is taken to be rigid,
$n_\mathrm P = 0$, and two bonds of the ligand are allowed to rotate,
namely, those connecting the amidine and the amino groups to the
benzene ring, $n_\mathrm L = 2$.  The published force field
\cite{amber7} does not provide a satisfactory torsion for the bond
between the benzene ring and the amidine group and this term was
determined using Gaussian 98 \cite{gaussian98} as $[-14.2
\cos(2 \phi) + 3.3 \cos(4 \phi) + 0.5 \cos(6 \phi)]\,\mathrm{kJ/mol}$,
where $\phi$ is the dihedral angle.

Five unbound and five bound canonical simulations of 1000 steps each
were carried out to find the portals.  The resulting configurations
were supplemented by those obtained by applying the symmetry operations
which leave the ligand invariant.  This gave 16 unbound and 8 bound
portals.  The configurations of the bound portals are all the same
``pose'' of the ligand on the protein and correspond to the symmetries
of $p$-amino-benzamidine given by rotating the amino, benzene, and
amidine groups by $180^\circ$.
The unbound ligand also exhibits 8 symmetries given by rotating the
amino and amidine groups by $180^\circ$ relative to the central benzene
ring and by including the mirror images.  However since the bond
parameters allow partially free rotation of the amidine group, each
symmetric set of configurations is represented by two portals.

We estimated a suitable value for $V_0$ of $0.39\times10^{-18}\,
\mathrm{m^3}$ based on the volumes and the mean energies of the computed
portals.  During the binding affinity calculation, wormhole moves were
attempted on 90\% of the steps (the other 10\% were standard Monte Carlo
moves).  Of these attempted wormhole moves, 97\% failed (inexpensively)
because the configuration was not in the chosen $w$.  Of the remaining
3\%, about 60\% lead to a successful move, of which about 40\% involved
switching from $\lambda =0$ to $1$ and \emph{vice versa}.  The major
computational cost in the free energy calculation is the evaluation of
the bound energies.  In this example, the wormhole moves required less
than 3 bound energy evaluations, on average, to effect the transition
from a bound configuration to an unbound one and back.  This enables an
accurate estimate to be made of the ratio of the averages in
Eq.~(\ref{kdcanon}) which after $5\times10^6$ steps yields $\mathrm pK_d
= 7.99 \pm 0.01$.  The error estimate is derived in the next section
and represents a 2\% relative error in $K_d$.

This example illustrates that the method is effective at allowing
sufficient transitions between the bound and unbound states to enable
the binding affinity of protein-ligand systems to be accurately
computed.  We note that our computed binding affinity differs from the
experimental results of $5.1$ to $5.2$
\nocite{maresguia65}\cite{maresguia65,schwarzl02}.
This discrepancy may be accounted for by modest, $\sim20\%$, errors in
the force field.

\section*{Error analysis}

In order to assess the errors in the computation of $\mathrm pK_d$ in
more detail, we carried out 10 independent runs similar to the one
described in previous section.  Each of these used the same portals
and the same value of $V_0$.  We computed cumulative estimates for
$\mathrm pK_d$ based on the first $s$ steps of the Markov chains.  When
forming the averages in Eq.~(\ref{kdcanon}) we sample every 100th step.
(As we shall see, there is a high degree of correlation within 100
steps; thus there would be little improvement in the estimate of
$\mathrm pK_d$ by sampling more frequently.)
\begin{figure}
\begin{center}
\vspace{-0.5ex}
\includegraphics[width=\figwidth]{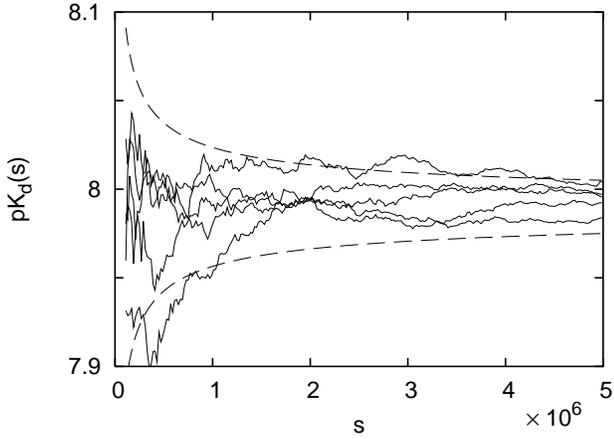}
\vspace{-1.5ex}
\end{center}
\caption{\label{cumulative}
Cumulative estimates $\mathrm pK_d(s)$ obtained by sampling every 100th
step from the first $s$ steps of 5 independent Monte Carlo runs.
The dashed lines shows
convergence as $1/\sqrt s$ to the mean value of $7.99$.}
\end{figure}
The results for 5 of these runs are shown in Fig.~\ref{cumulative}.  The
convergence toward the mean is as $1/\sqrt s$; after $5\times10^6$
steps, the error in $\mathrm pK_d$ has been reduced to about $\pm 0.01$.

For the purposes of further analysis, let us assume that the computation
is carried out with $E_\lambda^*(\Gamma) = E_\lambda(\Gamma)$ so that
all samples have the same statistical weight.  The probability that the
system is in the bound (resp.\ unbound) state is $p = \langle
\delta_{\lambda1} \rangle$ (resp.\ $q = 1-p = \langle \delta_{\lambda0}
\rangle$) and Eq.~(\ref{kdcanon}) becomes
\begin{equation}\label{kdcanona}
K_d = \frac1{V_0}\frac qp.
\end{equation}
The determination of $K_d$ is then
equivalent to estimating $q/p$ by taking the ratio of the
number of unbound and bound steps in the Monte Carlo simulation.  How
well this estimate converges depends, naturally, on the ``correlation
time'' of the system.  If wormhole moves rarely cause the system to
switch between bound and unbound states, the correlation time is large
and the convergence will be slow.

In order to make these ideas quantitative, we define the
$\lambda$-correlation function,
\[
C_t =
\langle (\lambda_s - p)
         (\lambda_{s+t} - p)
\rangle_s,
\]
where $\lambda_s$ is the value of $\lambda$ at simulation step $s$ and
$\langle\ldots\rangle_s$ denotes an average over steps.
\begin{figure}
\begin{center}
\vspace{-0.5ex}
\includegraphics[width=\figwidth]{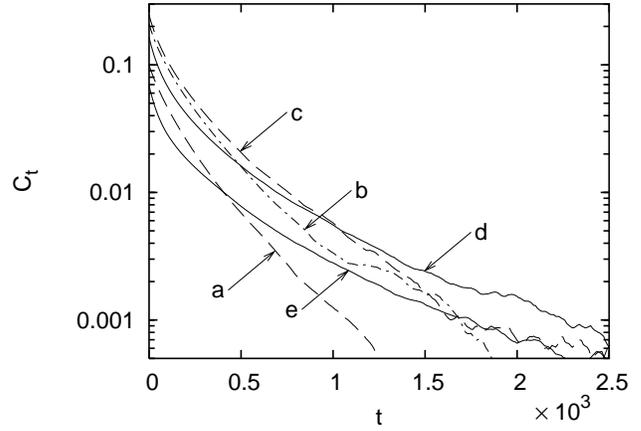}
\vspace{-1.5ex}
\end{center}
\caption{\label{correlation}
The $\lambda$-correlation function $C_t$.  The curves show $C_t$ for
$V_0/(0.39\times10^{-18} \,\mathrm{m^3}) =$ (a)~$\frac1{10}$,
(b)~$\frac13$, (c)~$1$, (d)~$3$, and (e)~$10$.
The ratio $q/p$ varies correspondingly (from
approximately $\frac18$ to $\frac{100}8$).  The correlation times are
(a)~$329$, (b)~$346$, (c)~$352$, (d)~$382$, and (e)~$415$.
In determining the
correlation times, we limit the sum in Eq.~(\ref{diffusion}) to $0< t
\le 2500$ in order to avoid including the small but noisy terms for
large~$t$.}
\end{figure}
Figure \ref{correlation} shows $C_t$ for several different values of
$V_0$.  (From Eq.~(\ref{kdcanona}), we have $q/p\propto V_0$.)  In a
Markov chain of length $s$, the expected number of bound states is $ps$,
while, for $s\rightarrow\infty$, the variance in the number of bound
states is $2Ds$, where
\begin{equation}\label{diffusion}
D = \frac12 C_0 + \sum_{t>0} C_t = \frac12 C_0 \tau.
\end{equation}
This provides us with the definition of the correlation time, $\tau$.
From Fig.~\ref{correlation}, we see that $C_t$ decays approximately
exponentially so that the sum converges.

We can compare the Monte Carlo simulation to a simpler Markov process of
independent trials, e.g., tossing a coin where the probability of heads
is $p$.  In this case, we have $C_{t>0} = 0$ and $D=\frac12 C_0 =
\frac12 pq$.  In the limit $s\rightarrow\infty$, the relative errors in
estimating $p$ for the Monte Carlo simulation will be the same as that
for $s/\tau$ tosses of the coin.  We can illustrate this by making 5
independent simulations of a coin tossing experiment to match the data
in Fig.~\ref{cumulative}, for which $p = 0.443$ and $\tau = 352$.
\begin{figure}
\begin{center}
\vspace{-0.5ex}
\includegraphics[width=\figwidth]{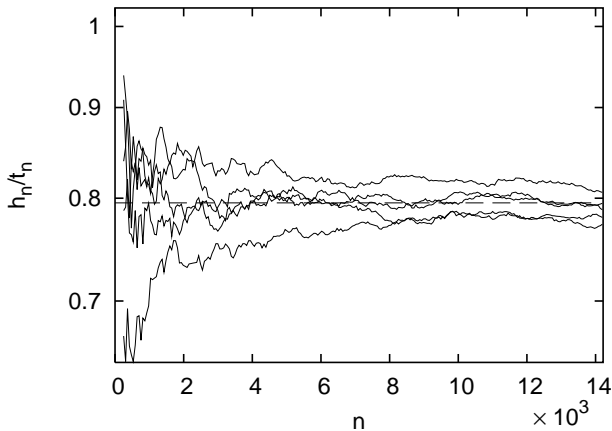}
\vspace{-1.5ex}
\end{center}
\caption{\label{cointoss}
Estimation of the bias of a coin with $p = 0.443$.  Five independent
runs are shown.  In each run, the coin was tossed $5\times 10^6/\tau =
14200$ times, where $\tau$ is the correlation time for
Fig.~\ref{correlation}(c), and the cumulative value of $h_n/t_n$ is
recorded.  The expected value, $p/q$, is shown as a dashed line.  The
axes have been adjusted to allow the figure to be directly compared with
Fig.~\ref{cumulative}.}
\end{figure}
The results are given in Fig.~\ref{cointoss}, where we plot $h_n/t_n$
against $n$ where $h_n$ (resp.\ $t_n$) is the number of heads (resp.\
tails) after $n$ throws.  As expected, Figs.~\ref{cumulative} and
\ref{cointoss} both exhibit the same convergence behavior.

Since the distribution of outcomes $h_n$ in the coin tossing experiment
is the binomial distribution, the mean and variance for $l_n =
\ln(h_n/t_n)$ can be calculated, yielding
\[
\langle l_n \rangle
  = \ln\biggl(\frac pq \biggr) +
  \frac1{2n}\biggl(\frac pq - \frac qp\biggr) +
  O(n^{-2}),
\]
\[
\langle (l_n - \langle l_n \rangle)^2 \rangle
  = \frac1{npq} + O(n^{-2}).
\]
The standard error in the estimate of $\mathrm pK_d$ from a Monte Carlo
run with $s$ steps is found by substituting $n=s/\tau$ and scaling by
$\ln 10$ (since $\mathrm pK_d$ is defined in terms of the common
logarithm) to give
\begin{equation}\label{err}
\Delta \mathrm pK_d \approx \frac1{\ln 10} \sqrt{\frac{\tau}{spq}}.
\end{equation}
In the case of the simulations shown in Fig.~\ref{cumulative}, we find
$\Delta \mathrm pK_d \approx 0.01$ consistent with the data in the
figure.  From Fig.~\ref{correlation}, we see that $\tau$ is weakly
dependent on $p$.  Thus for a given $s$, we minimize the error by
choosing $p = q = \frac12$.  Alternatively, we may wish to minimize the
error for a given amount of computational effort.  If bound steps are
$f$ times more expensive to carry out than unbound steps, the error is
minimized by taking $q/p = \sqrt f$, i.e., $p = 1/(1+\sqrt f)$.

\section*{Discussion}

The important aspects of this method that enable us to compute the
binding affinity of a ligand to a protein are: (1)~formulation in the
extended $[\Gamma,\lambda]$ space, which allows the unbound and bound
systems to be treated as a single canonical system and $K_d$ to be
expressed as the ratio of canonical averages, Eq.~(\ref{kdcanon});
(2)~the wormhole move which allows transitions between the bound and
unbound systems; and (3)~the use of an implicit solvation model which
reduces the number of degrees of freedom and so allows portals to be
identified.

A method similar to ours is ``simulated mutational equilibration,''
\cite{senderowitz97} which employs an extended phase space, uses an
implicit solvation model, and allows jumps between different systems.
However, in place of the wormhole move, this work employed a more
restrictive ``jumping between wells'' move, which limited its
applicability to computing the difference in the binding affinities for
two enantiomers.
In contrast, our wormhole method can be used to compute directly the
binding affinity of a ligand with several rotatable bonds to a protein
target.  This allows us to study a wide range of interesting drug-like
ligands.

In the special case of binding rigid molecules (for which we have
$E^*_0(\Gamma) = 0$), wormhole Monte Carlo is
isomorphic to a grand canonical Monte Carlo
simulation \cite{adams75}.  Consider a
grand canonical system with a single protein molecule and with ligand
molecules at a fixed chemical potential $\mu$; we make the additional
restriction that the ligand-ligand interaction energy is infinite, so that the
system can only accommodate 0 or 1 ligand molecule.  For the wormhole
simulation, we specify the bound portal to include the full simulation
volume with arbitrary orientation; the unbound port is degenerate and
corresponds merely to specifying $\lambda = 0$.  The  wormhole
moves from $\lambda = 0$ to $1$ and \emph{vice versa}
correspond to particle insertion and deletion in
the grand canonical simulation; and we can verify the acceptance
probabilities are the same.

The application described here can be extended by allowing a greater
degree of flexibility for the protein and the ligand.  This permits the
treatment of side-chain rotation and ligand-induced loop movement on the
part of the protein and the treatment of flexible rings for the ligand.
Many standard Monte Carlo techniques can be used with this method, if
appropriate---preferential sampling, early rejection, force bias, etc.
Wormhole moves could also be used to treat other discrete, or nearly
discrete, transitions.  Examples are: the treatment of molecules, such
as cyclohexane, which can assume distinct conformations; discrete sets
of side chain rotations in the protein; protonation and tautomerization
states for either molecule.  In each of these cases, the acceptance
probability for the transitions should account for the free energy
difference between the discrete states.

\section*{Acknowledgment}

This work was supported, in part, by the U.S. Army Medical Research and
Materiel Command under Contract No.\ DAMD17-03-C-0082.

\appendix
\ifpreprint
\subsection*{Appendix: Generalized Lambert projection}
\fi

The Lambert azimuthal equal-area projection projects a point on $S^2$ to
a point on a disk in $\mathbb R^2$.  Here, we will generalize this to
arbitrary dimensions, i.e., we will find a projection from $S^n$ to a
ball in $\mathbb R^n$.  (Thus the circle $S^1$ is projected into a
line segment; the surface of a sphere $S^2$ is project into a disk;
$S^3$ is projected into a sphere; etc.)  The projection is azimuthal, so
that directions from the pole are preserved in the projected space.

The sequence $S^n$ can be defined recursively by
\begin{eqnarray*}
S^0 &=& [\pm1], \\
S^n &=& [\cos\theta, \sin\theta \, S^{n-1}],
\end{eqnarray*}
where $\theta$ is the colatitude and $0 \le\theta\le\pi$.
The area of $S^n$ lying between $\theta$ and $\theta + d\theta$ is
therefore
\[
dA = a_{n-1} \sin^{n-1}\theta \,d\theta,
\]
where $a_n$ is the area of $S^n$.  We project that portion of $S^n$
with colatitude in $[0,\theta]$ to a ball in $\mathbb R^n$ of radius
$t$.  Equating the measures (area on $S^n$ and volume in $\mathbb R^n$),
we obtain
\[
v_n t^n = a_{n-1} \int_0^\theta \sin^{n-1}\theta'\,d\theta'
\]
where $v_n$ is the volume of a unit ball in $\mathbb R^n$.  Using the
relation,
\[
a_{n-1} = \left.\frac{d(v_n t^n)}{dt}\right|_{t=1} = n v_n,
\]
we obtain
\begin{equation}\label{genlambert}
t = \biggl(n\int_0^\theta \sin^{n-1}\theta'\,d\theta'\biggr)^{1/n}.
\end{equation}
The general ``equal-area'' projection is given by the
mapping
\[
[\cos\theta, \sin\theta \, S^{n-1}] \rightarrow \mathbf t,
\]
where $\mathbf t \in \mathbb R^n$, $\mathbf{\hat t}$ is given by the
point on $S^{n-1}$, and $t$ is given by Eq.~(\ref{genlambert}).  Some
special cases of Eq.~(\ref{genlambert}) are
\[
t = \left\{
\begin{array}{l@{\hspace{1em}}l}
\theta, & \mbox{for $n=1$,}\\
2\sin\frac12\theta, & \mbox{for $n=2$,}\\
\bigl[\frac34(2\theta - \sin2\theta)\bigr]^{1/3}, & \mbox{for $n=3$,}\\
2\sin\frac12\theta\bigl[\frac13(1+2\cos^2 \frac12\theta)\bigr]^{1/4},
& \mbox{for $n=4$.}\\
\end{array}
\right.
\]
The case $n=1$ corresponds to unwrapping a circle onto a line; and $n=2$
gives the Lambert azimuthal equal-area projection.  We are interested in
the case $n=3$ as a way of mapping orientations from unit quaternion
space to $\mathbb R^3$.  A quaternion $q = [\cos\theta,
\sin\theta \,\mathbf{\hat u}]$ represents a rotation of $\psi = 2\theta$
about $\mathbf{\hat u}$.  Noting that $q$ and $-q$ represent the same
rotation, we may make the restriction $0 \le \theta \le \frac12\pi$ .
We choose, therefore, to map this hemisphere of $S^3$ onto $\mathbf u$
in the unit ball, by making the substitutions $\theta = \frac12\psi$ and
$t = \bigl(\frac34\pi\bigr)^{1/3}u$ to give
\[
u = \biggl(\frac{\psi - \sin\psi}\pi\biggr)^{1/3}.
\]
(The other hemisphere, corresponding to $\frac12\pi<\theta\le\pi$, maps
onto the shell $1 < u \le 2^{1/3}$.)  We call this $\mathbf u$
representation of orientations ``turn space.''

\bibliography{free,wh}

\end{document}